\documentstyle[a4wide,12pt]{article}
\title{A reformulation of Hilbert's tenth problem through
Quantum Mechanics}
\author{Tien D Kieu~\footnote{kieu@swin.edu.au}\\
Centre for Atom Optics and Ultrafast Spectroscopy,\\Swinburne
University of Technology, Hawthorn 3122, Australia}
\begin{document}
\maketitle
\begin{abstract}
Inspired by Quantum Mechanics, we reformulate Hilbert's tenth
problem in the domain of integer arithmetics into either a problem involving a
set of infinitely coupled differential equations or a problem involving a Shr\"odinger
propagator with some appropriate kernel.
Either way, 
Mathematics and Physics could be combined for Hilbert's 
tenth problem and for the notion of effective computability.
\end{abstract}
\section{Introduction}
The twentieth century witnessed the remarkable discovery of the
limits of Mathematics, established within itself, through the noncomputable/undecidable results of Hilbert's
tenth problem, G\"odel's incompleteness theorem,
Turing halting problem, and their various extensions (see, for example,~\cite{davis, recursive}).
Such noncomputability and undecidability set the boundary for computation 
carried out by mechanical processes, and in 
doing so it help us to understand much better what can be so computed mathematically.

We have proposed elsewhere~\cite{kieu} a quantum algorithmic approach for the
non-computable Hilbert's tenth problem~\cite{davis, hilbert}, which is equivalent to the
Turing halting problem and intimately links to the concept of effective
computability as defined by the Church-Turing thesis.  While the proposal
is about some quantum processes to be implemented physically, it 
illustrates the surprisingly important r\^ole of Physics in the 
study of computability.  This is an unusual state
of affairs when Physics, which has its roots in the physical world out there,
could perhaps help setting the limits of Mathematics.

Inspired by this connection between the two, we present in this work
some mathematical reformulation of Hilbert's tenth problem.  
The reformulation is made possible since physical theories in general, and
Quantum Mechanics in particular, have enjoyed the support and rigour of
mathematical languages.  We wish to stress here that, despite of the
inspiration, the connection is established entirely in the domain
of Mathematics; we need not appeal to some real physical processes as we do 
with the proposed quantum algorithms in~\cite{kieu}.  It is hoped that such
reformulation may lead to new insights of the problem.

In the next section we briefly state the problem and the inspired
connection with operators acting on some infinite-dimensional Hilbert space.  
>From this we first derive a set of non-linearly, coupled differential equations (eqs.~(\ref{10}, \ref{4})) then also
a linear Scrh\"odinger equation (eq.~(\ref{Schroedinger})), each of which cases separately contains the 
sought-after decision result.  If one 
could find a universal procedure to derive certain information from these differential equations
(not necessarily by solving them explicitly but could be by other means) then one 
would have settled Hilbert's tenth problem in the positive.  

\section{Hilbert's tenth problem and Hilbert space}
At the turn of the last century, David Hilbert listed 25 important problems, among
which the problem number ten could be rephrased as:
\begin{quote}
\it
Given any polynomial equation with any number of unknowns and with integer
coefficients:  To devise a universal process according to which it can be
determined by a finite number of operations whether the equation has
integer solutions.
\end{quote}

This decision problem for such polynomial equations,
which are also known as Diophantine equations,
has eventually been shown in 1970 by Matiyasevich to be 
undecidable~\cite{hilbert,davis} in the Turing sense.  It is consequently
noncomputable/undecidable in the most general sense if one accepts, as 
almost everyone does, the Church-Turing thesis of computability.
Since exponential Diophantine, with the unknowns in the exponents
as in the example of the Fermat's last theorem, can be shown to be
Diophantine with supplementary equations, the study of Diophantine equations 
essentially covers the class of partial recursive functions, which is at 
the foundation of computability.  
The undecidability result is thus singularly important:  Hilbert's tenth
problem could be solved if and only if could be the Turing halting problem.
(See~\cite{ordkieu} also.)

Given a Diophantine equation with $K$ unknowns $x$'s
\begin{eqnarray}
D(x_1,\cdots,x_K) &=& 0,
\label{dio}
\end{eqnarray}
it suffices in general to consider the existence of nonnegative 
integer solutions.

Following~\cite{kieu} we link the equation above with the following
hermitean operator acting on some appropriate Fock space (a special type 
of Hilbert space)
\begin{eqnarray}
H_P &=& \left(D(a^\dagger_1 a_1,\cdots, a^\dagger_K a_K) \right)^2,
\label{hp}
\end{eqnarray}
where
\begin{eqnarray}
[a_j, a^\dagger_k] &=& \delta_{jk}, \nonumber\\ \nonumber
[a_k, a_j] &=& 0. \nonumber
\end{eqnarray}
The Fock space is built out of the ``vacuum" $|0_a\rangle$ by repeating 
applications of the creation operators $a^\dagger_j$.

The operator~(\ref{hp}) has a semidefinite and discrete spectrum 
$(D(n_1,\cdots,n_K))^2$.  This spectrum has an eigenstate
$|E_g\rangle$ corresponding to the smallest eigenvalue $E_g$.
(If the hermitean operator is 
considered as a hamiltonian for some dynamical process then these are 
the ground state and its energy, respectively.)

It is clear that the Diophantine equation~(\ref{dio}) has at least one
integer solution if and only if $E_g = 0$.

To sort out this $E_g$ among the infinitely many eigenvalues is almost 
an impossible
task.  The trick we will use, as inspired by quantum adiabatic processes,
is to tag the state $|E_g\rangle$ by some other known state $|E_I\rangle$ 
which is the ground state of some other operator $H_I$ and can be
smoothly connected to $|E_g\rangle$ through some continuous parameter 
$s\in[0,1]$.  To that end, we consider the interpolating operator 
\begin{eqnarray}
{\cal H}(s) &=& H_I + f(s)(H_P - H_I),\nonumber\\
&\equiv& H_I + f(s)W,
\label{1}
\end{eqnarray}
which has an eigenproblem at each instant $s$, 
\begin{eqnarray}
[{\cal H}(s) - E_q(s)]|E_q(s)\rangle = 0, &&q = 0, 1, \cdots
\label{eigen}
\label{2}
\end{eqnarray}
with the subscript ordering of the sizes of the eigenvalues, and $f(s)$ 
some continuous and monotonically increasing function in $[0,1]$
\begin{eqnarray}
f(0) = 0; && f(1) = 1.
\label{f}
\end{eqnarray}
Clearly, $E_0(0) = E_I$ and $E_0(1) = E_g$.
It turns out that for the function $E_0(s)$ to connect a ground state to 
another ground state we require that
\begin{eqnarray}
[H_P, H_I] &\not=& 0,
\label{symm}
\end{eqnarray}
and that $H_I$ has a discrete spectrum with a non-degenerate ground state
$|E_I\rangle$.  An example of such $H_I$ is, which we will employ from now on,
\begin{eqnarray}
H_I = \sum_{i=1}^K (a^\dagger_i -\alpha_i^*)(a_i -\alpha_i),
\label{HI}
\end{eqnarray}
in which case, $E_I = 0$ and $|E_I\rangle = |\alpha_1\cdots\alpha_K\rangle$,
that is, the coherent state
\begin{eqnarray}
|\alpha\rangle &=& {\rm e}^{-\frac{|\alpha|^2}{2}}\sum_{n=0}^{\infty}
\frac{\alpha^n}{\sqrt{n!}}|n_a\rangle.
\label{coherent}
\end{eqnarray}
where $\alpha\in {\cal C}$ and $|n_a\rangle$ are the eigenstates of $a^\dagger a$
with eigenvalues $n$.

\section{The spectral flow}
We now derive the differential equations for the tagging connection for 
the instantaneous eigenvalues and eigenvectors at different instant $s$
in~(\ref{eigen}). 

Note firstly that, from the normalisation condition $\langle E_q|E_q\rangle =
1$, we can write
\begin{eqnarray}
\langle E_q|\partial_s|E_q\rangle &=& -i\partial_s\phi_q,
\nonumber
\end{eqnarray}
for some real $\phi_q$.  This can be absorbed away with the redefinition
\begin{eqnarray}
{\rm e}^{i\phi_q(s)}|E_q(s)\rangle &\to& |E_q(s)\rangle,
\nonumber
\end{eqnarray}
upon which
\begin{eqnarray}
\langle E_q|\partial_s|E_q\rangle &=& 0.
\label{comp}
\end{eqnarray}
(The phases $\phi_q$ are related to Berry's phase in a dynamical process.)

Differentiating~(\ref{eigen}) with respect to $s$ yields
\begin{eqnarray}
[f'(s)W - \partial_s E_q] |E_q\rangle + [{\cal H} - E_q]\partial_s|E_q\rangle
&=& 0.
\label{3}
\end{eqnarray}
We next insert the resolution of unity at each instant $s$,
\begin{eqnarray}
{\bf 1} &=& \sum_{m=0}^\infty |E_m(s)\rangle \langle E_m(s)|,
\nonumber
\end{eqnarray}
just after ${\cal H}$ in~(\ref{3}) to get, vy virture of~(\ref{comp}),
\begin{eqnarray}
E_q\partial_s|E_q\rangle
&=& [f'(s)W - \partial_sE_q]|E_q\rangle +
\sum_{m\not = q}^\infty E_m \langle E_m|\partial_s|E_q\rangle
|E_m\rangle.
\label{7}
\end{eqnarray}
The inner product of the last equation with $|E_l\rangle$ gives
\begin{eqnarray}
(E_q-E_l)\langle E_l|\partial_s|E_q\rangle &=& f'(s)\langle E_l|W|E_q\rangle
-\partial_sE_q\delta_{ql}.
\label{8}
\end{eqnarray}
Thus, for $q\not=l$ this gives
the components of $\partial_s|E_q\rangle$ in $|E_l\rangle$, provided 
$E_q\not=E_l$ at any $s\in (0,1)$, a condition we will investigate in 
the next section.  Consequently, together with~(\ref{comp}),
\begin{eqnarray}
\partial_s|E_q\rangle &=& f'(s)\sum_{l\not = q}^\infty \frac{\langle
E_l|W|E_q\rangle}{E_q -E_l} |E_l\rangle.
\label{10}
\end{eqnarray}
Also, putting $q=l$ in~(\ref{8}) we have
\begin{eqnarray}
\partial_s E_q(s) &=& f'(s)\langle E_q(s)| W | E_q(s)\rangle.
\label{4}
\end{eqnarray}
Equations~(\ref{10}) and~(\ref{4}) form the set of infinitely coupled differential 
equations providing the tagging linkage we have been after.

Analytical and numerical methods
could now be employed to investigate the unknown ground state of $H_P$ from the 
constructively known spectrum of $H_I$ as the initial conditions.  In this reformulation, 
the Diophantine equation~(\ref{dio}) has at least one integer solution if and only if
\begin{eqnarray}
\lim_{s\to 1} E_0(s) &=& 0.
\label{answer}
\end{eqnarray}
The limiting process is necessary since $H_P$, i.e. ${\cal H}(1)$, will have a 
degenerate spectrum
because of certain symmetry ($H_P$ commutes with $a^\dagger_ia_i$).

The equations above are infinitely coupled and cannot be solved
explicitly in general.  But we are only interested in certain
information about the ground state.  And since the influence on the ground state 
by states having larger and larger indices diminishes more and more thanks to
the denominators in~(\ref{10}) (once no degeneracy is assured), 
this information may be derived, numerically or 
otherwise, with some truncation to a finite number of states involved.  
The size of the truncation cannot be universal and is of course dependent on 
the particular Diophantine equation under consideration.

\section{Expansion in the number basis}
We derive in this section an explicit form for the equations~(\ref{10}, \ref{4})
in the case of 2 unknowns, i.e. $K=2$ in~(\ref{dio}).  

At any instant $s$ we 
expand the eigenvectors $|E_q(s)\rangle$ in the basis consisting of the states $|n_1n_2\rangle$ of the
number operators $a^\dagger_1 a_1$ and $a^\dagger_2 a_2$,
\begin{eqnarray}
|E_q(s)\rangle &=& \sum_{n_1,n_2=0}C_{q; n_1n_2}(s)|n_1n_2\rangle,
\end{eqnarray}
upon which all the $s$-dependency are now carried by the coefficient
functions $C_{q; n_1n_2}(s)$ of the expansion.
Direct substitution of the expansion into~(\ref{10}, \ref{4}) gives
\begin{eqnarray}
\partial_s E_q(s) = f'(s)
\sum_{n_1,n_2}&&\left\{|C_{q_; n_1n_2}(s)|^2\left[D^2(n_1,n_2)- n_1-n_2-
|\alpha_1|^2 -|\alpha_2|^2\right]\right.\nonumber\\
&&+\left.\alpha_1\sqrt{n_1+1}C^{*}_{q;(n_1+1)n_2}(s)C_{q;n_1n_2}(s)  + {\rm c.c.}\right. \nonumber\\
&&+\left.\alpha_2\sqrt{n_2+1}C^{*}_{q;n_1(n_2+1)}(s)C_{q;n_1n_2}(s)+ {\rm c.c.}
\right\};
\label{eq1}
\end{eqnarray}
and
\begin{eqnarray}
\partial_s C_{q;n_1n_2}(s) = &&f'(s)\sum_{l\not=q}\frac{C_{l;n_1n_2}(s)}{E_q(s)-E_l(s)}\times\nonumber\\
&&\left\{
C^{*}_{l;n_1n_2}(s)C_{q;n_1n_2}(s)\left[D^2(n_1,n_2)- n_1-n_2-
|\alpha_1|^2 -|\alpha_2|^2\right]\right.\nonumber\\
&&+\left.\alpha_1\sqrt{n_1+1}C^{*}_{l;(n_1+1)n_2}(s)C_{q;n_1n_2}(s)\right.\nonumber\\
&&+\left.\alpha_1^*\sqrt{n_1+1}C^{*}_{l;n_1n_2}(s)C_{q;(n_1+1)n_2}(s)\right. \nonumber\\
&&+\left.\alpha_2\sqrt{n_2+1}C^{*}_{l;n_1(n_2+1)}(s)C_{q;n_1n_2}(s)\right. \nonumber\\ 
&&+\left.\alpha_2^*\sqrt{n_2+1}C^{*}_{l;n_1n_2}(s)C_{q;n_1(n_2+1)}(s)
\right\}
\label{eq2}
\end{eqnarray}

The appropriate initial conditions for these infinitely coupled differential equations 
can be derived from the eigenvalues and eigenvectors of the initial hermitean operator $H_I$.
If we choose to index the initial eigenvalues as
\begin{eqnarray}
(E_0(0), E_1(0), E_2(0),E_3(0), \cdots) &=& (0,1,1,2,\cdots),
\label{ini1}
\end{eqnarray}
then the coefficient functions at $s=0$ can be inferred as follows.
>From the expression for the coherent state~(\ref{coherent})
\begin{eqnarray}
|E_0(0)\rangle &=& |\alpha_1\alpha_2\rangle,\nonumber\\
&=& {\rm e}^{-\frac{|\alpha_1|^2+|\alpha_2|^2}{2}}\sum_{n_1,n_2=0}
\frac{\alpha_1^{n_1}\alpha_2^{n_2}}{\sqrt{n_1!n_2!}}|n_1n_2\rangle,
\end{eqnarray}
we arrive at
\begin{eqnarray}
C_{0;n_1n_2}(0) &=& {\rm e}^{-\frac{|\alpha_1|^2+|\alpha_2|^2}{2}} 
\frac{\alpha_1^{n_1}\alpha_2^{n_2}}{\sqrt{n_1!n_2!}}.
\label{ini2}
\end{eqnarray}

The next excited states are doubly degenerate, $E_{1}(0)=E_{2}(0)=1$,
\begin{eqnarray}
|E_{1}(0)\rangle &=& (a^\dagger_1 - \alpha_1^*)|\alpha_1\alpha_2\rangle,
\nonumber\\
&=& -\alpha_1^*|E_0(0)\rangle + {\rm e}^{-\frac{|\alpha_1|^2+|\alpha_2|^2}{2}}\sum_{n_1,n_2=0} 
\frac{\alpha_1^{n_1}\alpha_2^{n_2}}{\sqrt{n_1!n_2!}}\sqrt{n_1+1}|(n_1+1)n_2\rangle;
\end{eqnarray}
and
\begin{eqnarray}
|E_{2}(0)\rangle &=& (a^\dagger_2 - \alpha_2^*)|\alpha_1\alpha_2\rangle,
\nonumber\\
&=& -\alpha_2^*|E_0(0)\rangle + {\rm e}^{-\frac{|\alpha_1|^2+|\alpha_2|^2}{2}}\sum_{n_1,n_2=0} 
\frac{\alpha_1^{n_1}\alpha_2^{n_2}}{\sqrt{n_1!n_2!}}
\sqrt{n_2+1}|n_1(n_2+1)\rangle.
\end{eqnarray}
The last two equations subsequently yield
\begin{eqnarray}
C_{1;n_1n_2}(0) &=&  
\left\{
\begin{array}{ll}
-\alpha_1^*C_{0;n_1n_2}(0)+{\rm e}^{-\frac{|\alpha_1|^2+|\alpha_2|^2}{2}} 
\sqrt{n_1}\frac{\alpha_1^{(n_1-1)}\alpha_2^{n_2}}{\sqrt{(n_1-1)!n_2!}};&n_1\not=0;\\
-\alpha_1^*C_{0;0n_2}(0);&n_1=0.
\end{array}
\right.
\label{ini3}
\end{eqnarray}
And
\begin{eqnarray}
C_{2;n_1n_2}(0) &=&  
\left\{
\begin{array}{ll}
-\alpha_2^*C_{0;n_1n_2}(0)+{\rm e}^{-\frac{|\alpha_1|^2+|\alpha_2|^2}{2}} 
\sqrt{n_2}\frac{\alpha_1^{n_1}\alpha_2^{(n_2-1)}}{\sqrt{n_1!(n_2-1)!}};&n_2\not=0;\\
-\alpha_2^*C_{0;n_10}(0);&n_2=0.
\end{array}
\right.
\label{ini4}
\end{eqnarray}

The expressions~(\ref{ini1}, \ref{ini2}, \ref{ini3}, \ref{ini4}), {\em etc} $\ldots$ are
the initial conditions at $s=0$ for the differential equations~(\ref{eq1}, \ref{eq2}).  
However, because of the degeneracy in the initial conditions~(\ref{ini1}) we may have to
integrate the differential equations from some $s=\epsilon$ infinitesimally away from zero -- 
in which case the degeneracy should be lifted (as justified in the next section, there is no generic 
level crossing in the open interval $0<s<1$); and where the new initial conditions 
can be estimated numerically (to be very closed to the values at $s=0$)
by the familiar perturbation theory in quantum mechanics.

\section{No crossing for a single pair of levels}
We now argue that at any instant $s_0\in(0,1)$ if there is only one pair of
eigenvalues, say $(l,l+1)$ out of the infinitely many, can come very close 
together then they can never actually cross.  
The arguments are similar to 
those of perturbation theory for nearly degenerate levels~\cite{davydov}.

As can be seen from~(\ref{10}), in a neighbourhood
around $s_0$ we need only consider the two states $|E_l\rangle$ and
$|E_{l+1}\rangle$ as they are so strongly coupled that the rest can be
safely ignored.  As our world just becomes a two-dimensional space, we
can linearly decompose the two states at the next instant $(s_0+\delta s)$ 
in terms of the two at $s_0$.  Now with this decomposition, 
we need only to solve a two-by-two determinant for the eigenvalue problem for 
${\cal H}(s_0+\delta s)$.
The end result for the gap is
\begin{eqnarray}
\Delta_{l,l+1}(s_0+\delta s) &\equiv& E_{l+1}(s_0+\delta s) - E_l(s_0+\delta s),\nonumber\\
&=& \sqrt{[\Delta_{l,l+1}(s_0) + \delta sf'(s_0)(W_{l+1,l+1} - W_{l,l})]^2 +
4\delta s^2|f'(s_0)W_{l,l+1}|^2},\nonumber\\
\label{gap}
\end{eqnarray}
where $W_{i,j} = \langle E_i(s_0)|W|E_j(s_0) \rangle$.
The matrix element in the last term in the square root 
is just proportional to the difference 
\begin{eqnarray}
|\langle E_l(s_0)|{\cal H}(s_0+\delta s) - {\cal H}(s_0)|E_{l+1}(s_0)
\rangle|^2&=&
|\langle E_l(s_0)|{\cal H}(s_0+\delta s)|E_{l+1}(s_0) \rangle|^2. 
\label{diff}
\end{eqnarray}
Since
\begin{eqnarray}
[{\cal H}(s_0+\delta s), {\cal H}(s_0)] &=& \delta s f'(s_0)[H_P,H_I] 
+ O(\delta s^2)\not = 0,
\label{commutator}
\end{eqnarray}
{\em even restricting} to our two-dimensional subspace.  Then, as
the eigenvectors of ${\cal H}(s_0)$ cannot diagonalise
${\cal H}(s_0+\delta s)$, upon which the off diagonal elements~(\ref{diff}) cannot be zero.
Thus the gap~(\ref{gap}) can never be zero, no matter how vanishingly small it is at 
the previous instant.
(If the commutator~(\ref{commutator}) vanishes for our two-dimensional matrices, 
in contradiction to its nonvanishing for the operators in the whole Hilbert space, 
then our assumption of being able to ignore all other eigenstates apart from the two 
in consideration is not valid--perhaps because the gap at $s_0$ is not yet small enough.)

We note that from~(\ref{gap}) or~(\ref{4}) we can derive a differential equation 
for the gap and look for its minimum value.

The above arguments are only applicable for ``accidental degeneracy", that is 
when there is exactly one single pair of levels comes very closed.  The reasoning 
fails when we are not allowed to isolate a two-dimensional subspace and treat 
it separately as in the above.  Namely, it may fail when there are not 
two but three or more levels come crossing at one point; then the denominators
in~(\ref{10}) force us to consider a larger subspace.  The resulted determinant 
will have larger dimensions and not enough constraints to keep all the 
gaps non-zero.  

The above arguments for level avoidance also fail when there is another crossing pair, 
say $(q,q+1)$, 
elsewhere in the spectrum at {\it that same instant}.  Then the feedback
of that pair $(q,q+1)$ through their states to the right hand side of~(\ref{10}) 
for the pair $(l,l+1)$ may not be ignored; and we end up with a dimensionally 
larger subspace again.

But it is neither accidental nor typical that three or more levels cross
at exactly one point, or two or more pairs become degenerate at exactly the 
same instant.  Those events belong to a zero-measure set of events, unless
there must be a reason.  It should be a symmetry reason, 
that is, ${\cal H}(s_0)$ should commute with some other hermitean operator(s).  
Our mathematical elaboration above agrees with the observation of symmetry
and degeneracy in~\cite{adiabatic}.  

With care we can slightly modify the derivation
for~(\ref{10}, \ref{4}) to come up with similar equations even when there is some
degeneracy in $[0,1]$.  But for the condition~(\ref{answer}) to be the
indicator for the existence of solutions of the Diophantine equation, simple
topological consideration only requires
{\it that the initial ground state $|E_0(0)\rangle= |E_I\rangle$ is not degenerate and 
that this state doesnot cross with any other state in the open interval $s\in(0,1)$.}  
With the freedom of choice for $H_I$  satisfying~(\ref{symm}), we should be able to 
eliminate any symmetry in the open interval $s\in(0,1)$ for ${\cal H}(s)$ in order to 
have a stronger condition of totally avoided crossing.  This is because that
accidental symmetry cannot persist with widely different choices of $\alpha_i$ in
different starting $H_I$s (which result in different sets of differential equations that
can be used to cross confirm each other). 

Alternatively, one could also systematically eliminate the degeneracy effects, if any,
caused by some
accidental symmetry for a general Diophantine equation by considering a modified $H_P$
\begin{eqnarray}
H_P &\to& H_P + \sum_{i=1}^K (\epsilon_i a_i^\dagger + \epsilon_i^* a_i),
\label{modified}
\end{eqnarray}
with $|\epsilon_i|\ll 1$.  In the limits $|\epsilon_i|\to 0$, we recover our
original $H_P$ and would also be able to, by going through the limiting processes,
discover and eliminate any accidental symmetry.  It is clearly seen that the extra terms 
in~(\ref{modified}) will also remove any degeneracy of ${\cal H}(1) = H_P$.  (This is
similar to well-exploited physical trick in atomic physics of degeneracy removing by 
small biased external fields.)

\section{Hilbert's tenth and the Schr\"odinger equation}
The decision result for Hilbert's tenth problem is also contained in yet
another type of differential equation, apart from the nonlinear
equations~(\ref{10}, \ref{4}).  The linear equation is just the
Schr\"odinger equation which captures the dynamics of our quantum
algorithm~\cite{kieu}.  Let $|\psi(t)\rangle$ be the quantum state at time $t$,
its time evolution is given in quantum mechanics by the equation
\begin{eqnarray}
\partial_t |\psi(t)\rangle &=& -i{\cal H}(t/T)|\psi(t)\rangle,
\label{Schroedinger}\\
|\psi(0)\rangle &=& |\alpha_1 \cdots\alpha_K\rangle,\nonumber
\end{eqnarray}
where we have chosen the initial state at time $t=0$ to be the ground
state of $H_I$.

Unlike the case of nonlinear differential equations of a previous
section, now we could try to make use of a powerful computability result in analysis which is known as the First Main 
Theorem~\cite{pour-el}.  Essentially, the Theorem states that a {\it
bounded} linear operator from a Banach space to a Banach space which maps a computable
sequence of spanning vectors into another computable sequence will also map
{\em any} computable element into another computable
element. For the case at hand, our Schr\"odinger equation defines
a linear operator,
\begin{eqnarray}
U(T) &=& {\cal T} {\rm exp}\left\{ -iT\int_0^1{\cal H}(s) ds\right\},
\label{unitary}
\end{eqnarray}
where $\cal T$ is the time-ordering symbol, mapping the initial state to the final
state in the same separable Hilbert space.
Now,
our initial state $|\alpha_1 \cdots\alpha_K\rangle$
is computable by construction.  On the other hand, the linear operator~(\ref{unitary}) 
coming out of
the Schr\"odinger equation should be unitary and thus be bounded.  
Hence, the only remaining condition of the Theorem to be checked is
whether the image of a particular computable basis is computable or not.  
Speculatively, if it is the case then Hilbert's tenth problem
of integer arithmetic is decidable through the use of mathematical analysis tools
(plus those of the theory of infinite-dimensional operators)! 
This will be investigated elsewhere.

Nonetheless, we speculate that for the Schr\"odinger
equation to offer some new results here then there must be no level crossing
in the spectral flow.  But here as we start with the ground state of
$H_I$ we only require no level crossing for the instantaneous ground state,
unlike the situation with the nonlinear equations previously where we
have required no crossing for all levels.  Adapting Ruskai's arguments~\cite{ruskai}
we can show that the ground state of~(\ref{1}) is non-degenerate for $s\in(0,1)$.  
(The arguments are only applicable for the ground state but,
interestingly, the conclusion of the last section can also be supported
by them.  The trick is to use Ruskai's arguments for the
two-dimensional ``subspace" there; as there are only two levels, the lower
level is now the ground state.  Once again, this trick is not applicable
when we cannot, because of some symmetry reason as already discussed in the last section,
isolate such two-dimensional subspaces.  With more dimensions than two,
adaptation of Ruskai's cannot rule out the crossing of other states different from
the ground state.)

It should also be noted that the above quantum mechanical approach to 
Turing-noncomputable problems is in
contrast to the claim in~\cite{QTM} that quantum Turing machines
compute exactly the same class of functions, albeit 
perhaps more efficiently, which can be computed by classical Turing machines.  
However, the quantum Turing machine approach is a
direct generalisation of that of the classical Turing machines but with qubits
and some universal set of one-qubit and two-qubit unitary gates to build up,
step by step, dimensionally larger, but still dimensionally finite unitary operations.  
This universal set is chosen on its
ability to evaluate any desirable classical logic function.
Our approach, on the other hand, is from the start 
based on infinite-dimension Hamiltonians acting on some Fock space
and also based on the special properties and unique status of their ground states.  
The unitary operations are then the Schr\"odinger time evolutions.  
The infinite dimensionality together with the unique 
energetic status of the vacuum could be the reasons behind the 
ability to compute, in a finite number of steps, what the dimensionally
finite unitary operators of 
the standard quantum Turing computation cannot do in a finite number of steps. 
Note that it was the general Hamiltonian computation that was discussed 
by Benioff and Feynman~\cite{benioff, feynman} in the conception days of 
quantum computation.

Indeed, Nielsen~\cite{nielsen} has also found no logical contradiction
in applying the most general quantum mechanical principles to the
computation of the classical noncomputable, unless certain Hermitean
operators cannot somehow be realised as observables or certain unitary
processes cannot somehow be admitted as quantum dynamics.  And up to now
we do not have any evidence nor any principles that prohibit these kinds
of observables and dynamics.

\section{Concluding remarks}
Inspired by Quantum Mechanics, we have reformulated the question of
solution existence of a Diophantine equation into the question of certain
properties contained in an infinitely coupled set of differential equations.  
In words, we encode the answer of the former question into the smallest
eigenvalue and corresponding eigenvector of a hermitean operator 
whose integer-valued spectrum is bounded from below.
And to find these eigen-properties we next deform the
operator continuously to another operator whose spectrum is known.
Once the deformation is also expressible in the form of a set of nonlinearly coupled
differential equations, we could now start from the constructive
knowns as a handle to study the desired unknowns.

In addition, we also explicitly present a linear Schr\"odinger equation
whose solution at some time $T$ from an appropriate initial condition 
contains the information about the decision result for
the Diophantine equation under investigation.

Note that these reformulations are entirely based on mathematics. 
If a general mathematical method could be found to extract the required
information from these differential equations for any given Diophantine equation
then one would have the solution to Hilbert's tenth problem itself.  
This may be unlikely but not be as contradictory as it seems --because the 
unsolvability of Hilbert's tenth problem is only established in the 
framework of integer arithmetic and in Turing computability, not necessarily 
in Mathematics in  general.  Tarski~\cite{tarski}  has shown that the question 
about the existence of {\it real} solutions of polynomials over the reals is, 
in fact, {\it decidable}.

In the case of the linear Schr\"odinger~(\ref{Schroedinger}), we could also exploit
a powerful computability result in analysis which is known as the First Main 
Theorem~\cite{pour-el}.  With this Theorem, computability has been illustrated to
be indeed context/framework-dependent in the example of the classical wave equation: 
whether the solution to this equation is computable or not depends crucially on the initial 
functions and definitions of the norm employed.  We refer to the original literature for a
thorough discussion of this remarkable property.  
In order to establish the {\it mathematical} decidability of Hilbert's problem
in this particular reformulation, we will need to investigate (and this will be done elsewhere)
whether the unitary transformation~(\ref{unitary}) satisfies all the conditions of the Theorem
or not.

On the other hand, even though this Theorem for linear operators is not applicable to the set 
of non-linear equations~(\ref{10}, \ref{4}), such reformulation of
Hilbert's tenth problem with continuous variable has opened up many
new directions for further investigations.  

\section*{Acknowledgements}  I am indebted to Alan Head for discussions, 
comments and suggestions.  I would also like to thank John Markham 
and Andrew Rawlinson for discussions; Yuri Matiyasevich and Gabor Etesi 
for email correspondence.

\end{document}